\documentclass[apj,twocolappendix]{emulateapj_feii}
\usepackage{apjfonts}
\usepackage{hhline}
\usepackage{amsmath}
\usepackage{graphicx}
\usepackage{natbib}
\usepackage{epstopdf} %% for pdflatex
\usepackage{subfigure}
\usepackage{color}
\usepackage{comment}
\usepackage{longtable}

 \font\sevenrm=cmr7 scaled 1000
\newcommand{\CIV}{C {\sevenrm IV}}

\newcommand{\NV}{N~{\sevenrm V}}

\newcommand{\FeII}{Fe~{\sevenrm II}}
\newcommand{\MgI}{Mg~{\sevenrm I}}

\newcommand{\Hb}{H$\beta$}

\newcommand{\MgII}{Mg~{\sevenrm II}}
\begin{document}
\title{Strong correlation between \FeII/\MgII\ ratio and Eddington ratio of type 1 Active galactic nuclei}
\author{Jaejin Shin$^{1,2}$}
\author{Jong-Hak Woo$^{2}$\altaffilmark{,4}}
\author{Tohru Nagao$^{3}$}
\author{Minjin Kim$^{1}$}
\author{Hyeonguk Bahk$^{2}$}
\affil{
$^1$Department of Astronomy and Atmospheric Sciences, 
Kyungpook National University, Daegu 41566, Republic of Korea\\
$^2$Astronomy Program, Department of Physics and Astronomy, 
Seoul National University, Seoul, 08826, Republic of Korea \\
$^3$Research Center for Space and Cosmic Evolution, Ehime University, 
Bunkyo-cho 2-5, Matsuyama, Ehime 790-8577, Japan
}

\altaffiltext{4}{Author to whom any correspondence should be addressed}

\begin{abstract}
The \FeII/\MgII\ line flux ratio has been used as an indicator of the Fe/Mg abundance ratio in
the broad line region (BLR) of active galactic nuclei (AGNs). 
On the basis of archival rest-frame UV spectra obtained via the {\it Hubble Space Telescope} and 
the Sloan Digital Sky Survey, we investigate the \FeII/\MgII\ ratios of type 1 AGNs at z < 2.
Over wide dynamic ranges of AGN properties (i.e., black hole mass, AGN luminosity, and Eddington ratio),
we confirm that the \FeII/\MgII\ ratio strongly correlates with Eddington ratio but not with black hole mass, 
AGN luminosity, or redshift. Our results suggest that the { metallicity} 
in the BLR are physically related to the accretion activity of AGNs, but not to the global properties of galaxies 
(i.e., galaxy mass and luminosity). With regard to the relation between 
{ the BLR metallicity} and the accretion rate of AGNs, we discuss that metal cooling may play an important 
role in enhancing the gas inflow into the central region of host galaxies, resulting in the high accretion rate of AGNs. \\
\end{abstract}
\keywords{
     galaxies: active ---
     galaxies: ISM ---
     galaxies: nuclei ---
     quasars: emission lines ---
     ultraviolet: galaxies
}

\section{INTRODUCTION} \label{section:intro}
In understanding galaxy evolution, the chemical properties of galaxies have been considered 
as one of the key parameters since they are closely related to the star formation history 
(see \citealt{Maiolino2019} for a review).
For normal galaxies, gas metallicity exhibits tight scaling relations with 
galaxy properties (i.e., mass and luminosity) with these relations evolving 
as a function of redshift up to $\sim$3.5
\citep[e.g.,][]{Erb2006,Maiolino2008,Mannucci2009,Yabe2012,Onodera2016},
which indicates the cosmic evolution of galaxy metallicity. 
At a higher redshift (i.e., z > 3.5), however, 
the characteristic of galaxy metallicity is poorly understood 
since it is challenging to measure the gas metallicity of normal galaxies
at such redshift levels \citep[but see e.g.,][]{Nagao2012,Shapley2017,Tadaki2019}.

Alternatively, active galactic nuclei (AGNs) can be investigated thanks to their high luminosity.
Through theoretical models (e.g., CLOUDY), several ultraviolet (UV) line flux ratios 
(e.g., N {\scriptsize V}$\lambda$1240/C {\scriptsize IV}$\lambda$1549; hereafter \NV/\CIV) 
have been proposed as indicators of the metallicity of the broad line region (BLR) of AGNs 
with these line ratios have exhibiting a correlation with AGN luminosity \citep[e.g.,][]{Hamann1992,Hamann1993,Nagao2006}. 
Since AGN luminosity is determined by the combination of the mass of supermassive black holes (SMBHs)
and Eddington ratio, there were efforts to reveal the physical origin of the BLR metallicity -- AGN luminosity 
relation between black hole mass and Eddington ratio. 

On the one hand, it has been found that BLR metallicity correlates with black hole mass \citep{Warner2004, Matsuoka2011b}. 
The mass of SMBHs shows scaling relations with the properties of their host galaxies 
(i.e., mass, luminosity and stellar velocity dispersion), implying the co-evolution of SMBHs and their host galaxies 
\citep[][]{Magorrian1998,Ferrarese2000,Gebhardt2000,Marconi2003,Haring2004,Kormendy2013,Woo2013,Woo2015}. 
Therefore, based on the relation between the global properties (i.e., mass) and the gas metallicity of galaxies, 
BLR metallicity can be expected to be connected to the galaxy metallicity. 
On the other hand, \cite{Shemmer2004} reported a correlation between BLR metallicity and Eddington ratio, but not black hole mass.  
This finding was confirmed by \cite{Shin2013}, who investigated low redshift AGNs over a larger dynamic range  (e.g., $\sim$3 dex) 
of black hole mass and Eddington ratio. Based on the findings, they discussed a physical connection between
the accretion activity of AGNs and the star formation history of the central region of galaxies.

Interestingly, no cosmic evolution of the BLR metallicity based on \NV/\CIV\ has been reported up to redshift$\sim$6 
 \citep[e.g.,][]{Nagao2006,Juarez2009}, suggesting that the BLR metallicity was already enriched 
at a universe age of 1 Gyr (z$\sim$6) and that the first star formation episode occurred at a redshift of 
higher than $\sim$10. This is inconsistent with the trend of galaxy metallicity, which clearly exhibits cosmic evolution 
\citep[e.g.,][]{Maiolino2008,Troncoso2014}.

In understanding the chemical properties of the BLR, another flux ratio between UV \FeII\ multiplet and 
\MgII$\lambda$2800 (hereafter \FeII/\MgII), has been investigated as a first-order proxy of the Fe/Mg abundance ratio in the BLR. 
Similar to the BLR metallicity indicator (i.e., \NV/\CIV), the \FeII/\MgII\ ratio also shows no dependency on redshift 
up to z$\sim$7 \citep[e.g.,][]{Barth2003,Maiolino2003,DeRosa2011,Mazzucchelli2017,Shin2019,Onoue2020,
Sameshima2020,Schindler2020}, which supports the idea of an absence of cosmic evolution in the chemical properties of the BLR. 
However, in contrast to the BLR metallicity indicators, the \FeII/\MgII\ ratio shows different relations with AGN properties since the 
\FeII/\MgII\ ratio correlates with Eddington ratio and black hole mass, but not with AGN luminosity 
\citep[][see also \citealt{Sameshima2017}]{Dong2011,Shin2019}.

In our previous study \citep[][hereafter Paper I]{Shin2019}, we discussed potential sample selection bias
that likely led to the absence of a correlation between the \FeII/\MgII\ ratio and AGN luminosity. 
The limitation of the previous studies (i.e., \citealt{Dong2011} and Paper I) is that only luminous 
and massive AGNs were investigated (i.e., log L$_{\rm Bol}>$46, M$_{\rm BH}$ > $10^{8}\rm M_{\odot}$). 
The enrichment of galaxy metallicity is found to be less significant in massive galaxies as a function of cosmic time, 
which may mean that massive galaxies are likely chemically matured \citep[see e.g.,][]{Maiolino2008}. 
Similarly, the luminous and massive AGNs can be likely chemically matured. This, in turn, may 
be a reason why no dependency of the \FeII/\MgII\ ratio on redshift and AGN luminosity was observed
in the previous works. 

Motivated by this consideration, we investigate the \FeII/\MgII\ ratio of type 1 AGNs at z < 2 over wide dynamic ranges 
of AGN properties (i.e., black hole mass, AGN luminosity, and Eddington ratio) to better understand the relation between
the \FeII/\MgII\ ratio and the AGN properties. 
In section 2, we describe the sample and data. The analysis is described in section 3, 
and our main results and discussion are presented in sections 4 and 5. We adopt a cosmology of 
$H_{\rm 0}= 70$ km  s$^{-1}$ Mpc$^{-1}$, $\Omega_{\Lambda}=0.7$ and $\Omega_{\rm m}=0.3$.  \\

\section{Sample and Data} \label{section:intro}
In this work, we investigate two type 1 AGN samples: (i) low redshift type 1 AGNs (z < $\sim$0.4)
with available archival UV spectra and (ii) Sloan Digital Sky Survey (SDSS) quasars at 0.46 < z < 2.02. 

For the low redshift AGNs, we first selected 84,040 type 1 AGNs at z < 0.75 from the
Million Quasars (MILLIQUAS) catalog (Version 6.3, June 2019, \citealt{Flesch2015}), with 
an object classification of either A (25,989) or Q (58,051). Among them, we found 77 objects 
with available archival Hubble Space Telescope (HST) spectroscopic data 
(i.e., COS, STIS, and FOS), which was searched within a radius of 3\arcsec, 
covering the wavelength range of 2600-3050\AA\ in the rest frame. 
In previous works of \FeII/\MgII\ ratio, \FeII\ flux has been generally fitted and integrated over the wavelength 
range of 2200-3090\AA\ (e.g., \citealt{Maiolino2003,Sameshima2017}; Paper I). However, the requirement of the 
wide wavelength range limits sample size. In Paper I, we demonstrated that a narrower wavelength range of 
2600--3050\AA\ is acceptable for the flux estimation of \FeII, albeit a non-negligible uncertainty could emerge. 
This wavelength range of 2600-3050 was thus adopted to expand the sample size.

We retrieved the single epoch spectra of the 77 objects that showed the best data quality (i.e., signal-to-noise ratio, 
SNR, at 3000\AA\ continuum).
{ We then finalized the sample of 29 type 1 AGNs using the following five criteria,
(i) amplitude-to-noise ratio of \MgII, ANR$_{\rm MgII}$  > 5, (ii) ANR of \FeII, ANR$_{\rm FeII}$  > 3,
(iii) SNR at 3000\AA\ continuum, SNR$_{3000}$ >3, 
(iv) no narrow or broad absorption lines (NAL or BAL) objects, and (v) no type 2 AGNs.
The first, second, and third criteria were adopted to obtain reliable measurements of \MgII\ and \FeII\ fluxes and AGN luminosity. 
For the second criteria, we defined the ANR$_{\rm FeII}$ as the ratio of the peak flux density of the best-fit iron model to the noise at 3000\AA.
There is one issue for measuring the ANR$_{\rm FeII}$ that \FeII\ is blended with \MgII, the power-law continuum, and Balmer pseudo continuum.
This means that deblending \FeII\ is required through multi-component fitting to measure the ANR$_{\rm FeII}$.
Moreover, due to the blending with \MgII, the choice of \MgII\ line profile in the multi-component fitting can affect the value of the ANR$_{\rm FeII}$. 
In this work, we adopted the ANR$_{\rm FeII}$ from the fitting with double Gaussian profile for \MgII. We refer to Section 3 for more details.}
We excluded three AGNs with prominent absorption features (NAL: NGC 4151 and NGC 3227, BAL: UGC 8058)
since the absorption features can severely affect the line flux estimations of both \MgII\ and \FeII. 
Lastly, we found three type 2 AGNs (Mrk 348, Mrk 3, and NGC 1068), which are misclassified as type 1 AGNs in the MILLIQUAS catalog. 
Since type 2 AGNs do not present broad emission lines (i.e., \MgII\ and \FeII), which is our main area of
interest, they were excluded. The information on the final sample is presented in Table~1.
We note that 
{ 11 of the 29 type 1 AGNs ($\sim$38\%)} 
cover the full wavelength range of \FeII, 2200-3090\AA.

In addition, we included a number of SDSS quasars from the SDSS DR7 quasars catalog \citep{Shen2011}.
We selected 5,441 quasars in the redshift range of 0.46 < z < 2.02, whose spectrum covers the rest-frame wavelength range of 2600-3050\AA,
and the SNR at 3000\AA\ continuum is larger than 20. 
 { Among them, we excluded 1,380 objects with strong absorption lines 
(i.e., BAL and NAL) and/or weak emission lines (i.e., ANR$_{\rm \MgII}$ < 5 and ANR$_{\rm \FeII}$ < 3), leaving a final sample of 4,061 quasars. 
Note that this sample includes the 3,164 quasars in a narrower redshift range (i.e., 0.75 < z < 1.96), 
which were investigated in Paper I, and the majority (895) of the newly included 897 quasars are at low redshift (i.e., 0.46<z<0.75).}\\

%%%%%%%%%%%%%%%%%%%%%%%%%%%%%%%%%%%%%%
%%%%%%%%%%%%%%%Table1%%%%%%%%%%%%%%%%%
%%%%%%%%%%%%%%%%%%%%%%%%%%%%%%%%%%%
%%%%%%%%%%%%%%%%%%%%%%%%%%%%%%%%%

\begin{deluxetable*}{lcccl} 
\tablewidth{0pt}
\tablecolumns{5}
\tabletypesize{\scriptsize}
\tablecaption{Log of archival HST UV spectroscopic data}

\tablehead{
\colhead{Object} &
\colhead{Redshift} &
\colhead{Instrument} &
\colhead{Grating} &
\colhead{Obs. Date} 
\\
\colhead{(1)} &
\colhead{(2)} &
\colhead{(3)} &
\colhead{(4)} &
\colhead{(5)} 
}

\startdata
                  KUG 0003+199 	&$	0.026	$&	FOS	&	G270H	&	1994	Dec	16	\\
              FIRST J0012-1022 	&$	0.228	$&	STIS	&	G430L	&	2011	May	14	\\
                     PGC 87392 	&$	0.029	$&	FOS	&	G270H	&	1996	Jul	30	\\
                    ESO 113-45 	&$	0.046	$&	FOS	&	G270H	&	1993	Jan	22	\\
                  HE 0132-4313 	&$	0.237	$&	FOS	&	G270H, G400H	&	1996	Sep	21	\\
                       3C 48.0 	&$	0.367	$&	STIS	&	G430L	&	2011	Mar	3	\\
      SDSS J015530.02-085704.0 	&$	0.165	$&	STIS	&	G430L	&	2011	Feb	2	\\
                       AKN 120 	&$	0.033	$&	FOS	&	G270H	&	1995	Jul	29	\\
                   PG 0844+349 	&$	0.064	$&	FOS	&	G270H	&	1992	Jan	10	\\
                  KUG 0921+525 	&$	0.035	$&	STIS	&	G230L	&	2017	Dec	25	\\
      SDSS J094603.94+013923.6 	&$	0.220	$&	STIS	&	G430L	&	2011	Mar	24	\\
                      NGC 3516 	&$	0.009	$&	FOS	&	G270H	&	1996	Feb	21	\\
                 MCG 10-16-111 	&$	0.028	$&	STIS	&	G230L	&	2013	Jul	12	\\
                  MCG 9-19-073 	&$	0.021	$&	STIS	&	G230L	&	2013	Apr	29	\\
                    CGCG 13-24 	&$	0.020	$&	STIS	&	G230L	&	2013	Jun	7	\\
MARK 50	&$	0.024	$&	STIS	&	G230L	&	2012	Feb	12	\\
                        3C 273 	&$	0.158	$&	STIS	&	G430L	&	1999	Jan	31	\\
               RXS J12517+2404 	&$	0.188	$&	STIS	&	G430L	&	2011	Apr	26	\\
                      3C 277.1 	&$	0.320	$&	FOS	&	G400H	&	1991	Nov	2	\\
NGC 5548	&$	0.017	$&	FOS	&	G270H	&	1992	Jul	5	\\
                     UGC 10120 	&$	0.031	$&	STIS	&	G230L	&	2017	Aug	29	\\
                  SBS 1701+610 	&$	0.165	$&	STIS	&	G430L	&	1999	May	2	\\
      SDSS J171448.50+332738.3 	&$	0.181	$&	STIS	&	G430L	&	2010	Oct	15	\\
                   CGCG 229-15 	&$	0.027	$&	STIS	&	G230L	&	2013	Jul	23	\\
NGC 6814	&$	0.006	$&	STIS	&	G230L	&	2013	May	7	\\
                      MARK 509 	&$	0.035	$&	STIS	&	G230L	&	2017	Oct	22	\\
                   B2 2201+31A 	&$	0.298	$&	FOS	&	G270H, G400H	&	1991	Sep	6	\\
                   MR 2251-178 	&$	0.064	$&	FOS	&	G270H	&	1996	Aug	2	\\
                      NGC 7469 	&$	0.017	$&	FOS	&	G270H	&	1996	Jun	18	
\enddata
\label{table:prop}

\tablecomments{
    Col. (1): Target name from the MILLIQUAS catalog \citep{Flesch2015}.   }
\end{deluxetable*}

\section{Analysis} \label{section:intro}
{ To measure the \FeII/\MgII\ ratios of the type 1 AGNs, we performed multi-component 
fitting analysis using the components of \MgII, \FeII, the power-law continuum and Balmer pseudo 
continuum, as performed in Paper I. For the fitting of \FeII, we adopted the iron template provided 
by \cite{Tsuzuki2006}. Note that we did not use the iron template of \cite{Vestergaard2001} because 
of the lack of iron information around \MgII\ (2770-2820\AA), which may lead over- or under- estimation 
of \MgII\ and \FeII\ fluxes, respectively (\citealt{Dietrich2003,Woo2018}; Paper I; \citealt{Schindler2020}).
In addition to the analysis method described in Paper I, we considered two more issues, 1) the line profile of \MgII,
and 2) the contribution of host galaxy in UV spectra, which will be described as follows.\\

\subsection{Line profile of \MgII}
In previous works, \MgII\ has been analyzed with various line profiles because of the complex line profile
of \MgII. Among them, double Gaussian and Gauss-Hermite series have been widely adopted to reproduce 
the observed \MgII\ line profile showing a wing component \citep[e.g.,][]{Marziani2013, Kovacevic2015,Woo2018,
Bahk2019,Le2019,Popovic2019,Le2020}. 
On the other hand, for AGNs with high accretion rate, it has been reported that a Lorentzian profile
gives better fitting result than the double Gaussian and Gaussian profiles \citep{Marziani2013}. 

In Paper I, we adopted a single Gaussian profile for \MgII\ fitting due to the low data quality of high redshift 
quasars (i.e., at z$\sim$3) in the work. As a comparison, we performed another fitting with a 6th 
order Gauss-Hermite profile for \MgII\ and found that the measured \FeII/\MgII\ ratios from fitting with
the single Gaussian and Gauss-Hermite profiles are consistent within $\sim$7\%. 
However, we did not test fitting with other line profiles (i.e., double Gaussian and Lorentzian).

To better understand whether the \MgII\ line profiles can affect the measurements the \FeII/\MgII\ ratio, 
we adopted various line profiles for \MgII, 1) single Gaussian, 
2) double Gaussian, 3) 6th order Gauss-Hermite series, and 4) Lorentzian profiles in the multi-component fitting analysis.

\subsection{Host galaxy continuum}
In AGN spectra, there could be non-negligible contamination from their host galaxy. 
With the detection of several stellar absorption lines (i.e., Mgb, Fe5270, \& Fe5335),
the host galaxy contamination in optical window has been identified and removed for the reliable 
investigation of AGN properties \cite[e.g.,][]{Park2012a,Park2015, Bon2020}. \cite{Shen2011}
statistically found that low luminous AGNs ($\rm log L_{\rm 5100\AA}$ < 45 or $\rm log L_{\rm Bol}$ < 46) are more likely 
contaminated by the host galaxy continuum (e.g., $\sim$20\%\ for AGNs with $\rm log L_{\rm 5100\AA}$ $\sim$ 44.5 or 
$\rm log L_{\rm Bol}$ $\sim$ 45.5). Moreover, \cite{Bon2020} reported that host galaxy continuum could affect the shape of
\FeII\ in the optical window. Similarly, the host galaxy contamination can be expected at UV, especially due to
young stellar population \citep[e.g.,][]{Leitherer2011}, which could affect the measurement of \FeII\ flux and AGN luminosity 
in the UV window. 

{
However, it seems challenging to investigate the host galaxy contamination in this work due to the following reason. 
In the wavelength of 2600--3050\AA, strong absorption lines are \FeII\ $\lambda$2600, \MgII, and
\MgI\ $\lambda$2852, while we have no AGNs exhibiting prominent absorption features at those lines in our sample of 
low redshift AGNs and SDSS quasars. During our sample selection process, we detected such AGNs 
(e.g., NGC 4151 an NGC 3227, see Section 2). However, we discarded them based on the assumption that they are NAL AGNs. 
We note that 15 targets out of our 29 low redshift AGNs exhibit prominent absorption line features, which are not
related to \FeII\ $\lambda$2600, \MgII, and \MgI\ $\lambda$2852. By considering the redshift of the 15 targets, 
all of their absorption lines are likely Milky way absorption lines \citep{Savage2000}, 
which are produced by gas in the disk and halo of the Milky Way. Therefore, the absorption lines 
are not related to the host galaxy continuum.}
Since we can not properly investigate the host galaxy contamination without 
stellar absorption lines, we decided not to consider the host galaxy contamination in this work.
We note that AGNs with the nondetection of \MgII\ absorption line suggest that 
their host galaxy contamination maybe not significant in their UV spectra. 
To investigate the host galaxy contamination in UV, one can use high-quality spectra covering
wide wavelength range (i.e., UV to optical) and showing prominent absorption lines (Mgb, Fe5270, and Fe5335) 
in future.\\

\subsection{Fitting procedure and measurement}
With the four line profiles for \MgII, we performed the multicomponent fitting analysis.
We note that we accepted the fitting with the double Gaussian profile only if the 
ANR of each Gaussian component is all larger than three. 
Otherwise, we adopted the fitting with the single Gaussian line profile.

During the fitting process, we did not give any kinematic constraints 
between \MgII\ and \FeII\ to avoid any possible bias. This is different from
Paper I, which gave constraints that \MgII\ and \FeII\ have the same velocity shift
and similar velocity dispersion ($\Delta \sigma$ < 10\%).
However, we note that the kinematic constraint does not significantly affect the 
measurements of \MgII\ and \FeII\ fluxes and the \FeII/\MgII\ ratio as discussed in Paper I (see their Figure~8).

From the best-fit model, we measured the flux and velocity dispersion of \MgII, whereas
for the flux of \FeII, we integrated the best iron model over the wavelength range, 2600--3050\AA.
In Figure~1, we present the comparisons of the \MgII\ flux, \FeII\ flux, and \FeII/\MgII\ ratio, which were 
measured from the fitting with the four \MgII\ line profiles. 
From the comparisons, we found three trends. First, the measurements from the fitting with the double Gaussian and
6th order Gauss-Hermite line profiles are consistent. Second, the fitting with double Gaussian line profile
gives larger \MgII\ fluxes but smaller \FeII\ fluxes and \FeII/\MgII\ ratios than those from the single Gaussian fitting. 
This would be because the wing component of \MgII\ can not be fitted with the single Gaussian profile.
Third, the double Gaussian fitting provides smaller \MgII\ flux and larger \FeII\ flux and \FeII/\MgII\ ratio than
the Lorentzian fitting, suggesting that the Lorentzian profile tends to more fit the wing component of MgII line than the double Gaussian.
While their are non-negligible scatters and offsets described above, the measurements from the fitting with various \MgII\ line profiles,
generally show strong linear relationships between each other. }

 \begin{figure*}{}
\includegraphics[width = 0.94\textwidth]{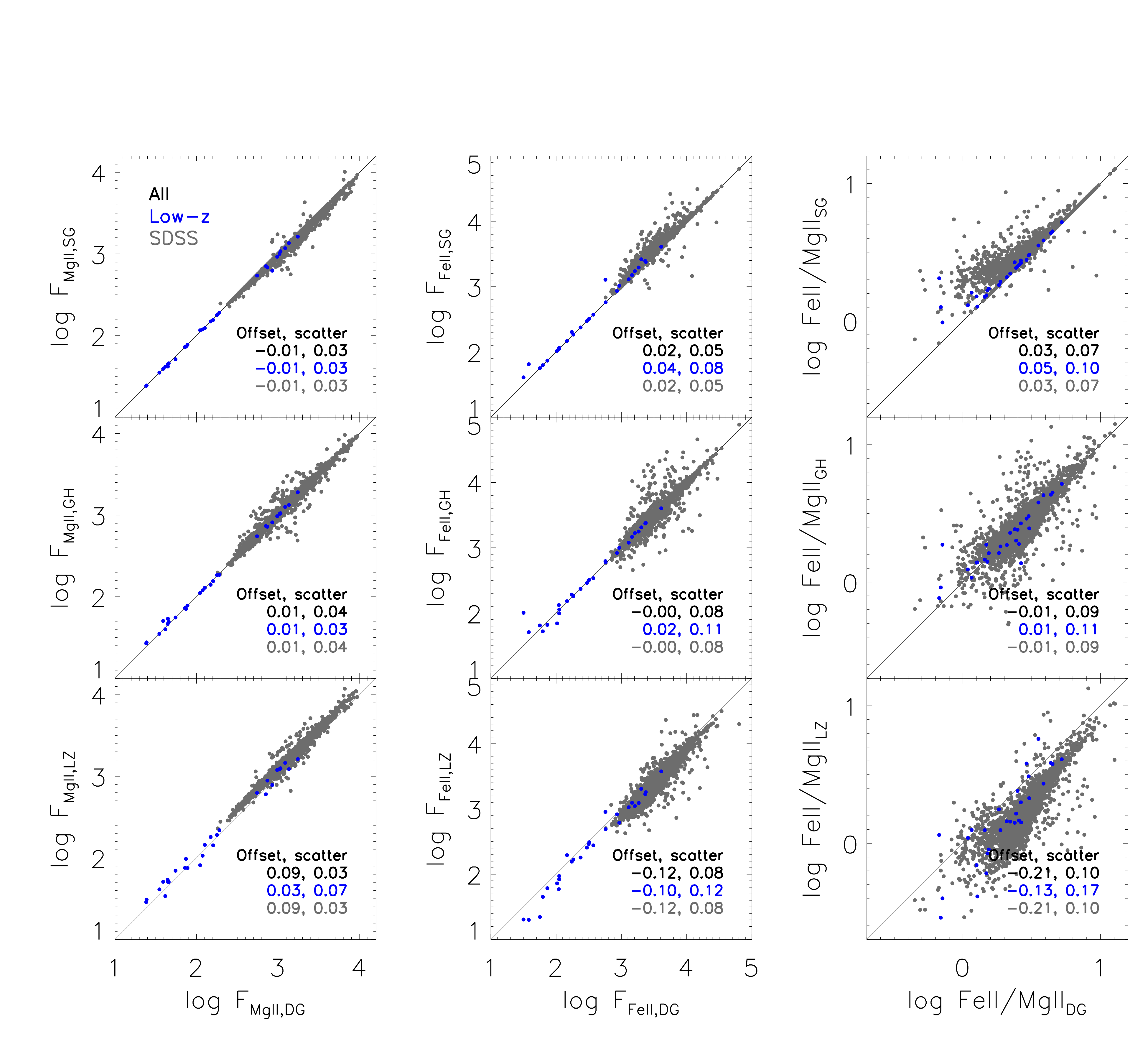}
\caption{
{ Comparison of \FeII\ flux (left), \MgII\ flux (center), and the \FeII/\MgII\ ratio (right) 
measured from fitting with various line profiles for \MgII. X-axis shows the measurement from 
double Gaussian line profiles and Y-axis shows the measurements from other profiles 
(top: single Gaussian, middle: 6th order Gauss-Hermite, and bottom: Lorentzian). 
Blue and gray represent the sample of low redshift AGNs and SDSS quasars, 
respectively. The offset and scatter given in each panel are the average and the standard deviation
of the difference between the two measurements in each panel.}\\
\label{fig:allspec1}}
\end{figure*}

We also estimated the AGN properties (i.e., black hole mass, AGN luminosity, and Eddington ratio) 
using the velocity dispersion of \MgII\ and the monochromatic luminosity at 3000\AA\ continuum. 
We calculated black hole mass using the equation given in \citet[][see also \citealt{Woo2018,Bahk2019}]{Le2020} 
with the coefficients of their scheme 2. 
{ Note that the \MgII-based black hole mass measurement 
has an uncertainty of 0.4-0.7 dex due to various issues
\citep[i.e., the virial factor and AGN variability, see e.g.,][and references therein]{Le2020}.}
The bolometric luminosity was converted from the 3000\AA\ monochromatic 
luminosity with a bolometric correction factor of 5.15 \citep{Shen2008}. 
Lastly, we calculated the Eddington ratio by dividing the bolometric luminosity by the Eddington luminosity, 
which is determined by black hole mass. We also estimated the uncertainties of the line fluxes and 
AGN properties using Monte-Carlo simulations
with 1000 mock spectra created from randomizing flux within 1$\sigma$ error. 
We list the properties of emission lines and AGN with their uncertainties in Table~2.

{ We also checked whether the choice of the \MgII\ line profile affects the estimation of AGN properties. 
First of all, we found that AGN luminosities from fitting with the four different line profiles are consistent within $\sim$0.01 dex. 
Secondly, we obtained consistent black hole mass from the fitting with the line profiles of single Gaussian, double Gaussian, and 
6th order Gauss-Hermite series within $\sim$ 0.1 dex, while they are significantly lower ($\sim$20 times) than that from the Lorentzian fitting. 
Due to the consistent AGN luminosities, the difference in black hole mass is mainly from the different velocity dispersion of \MgII, 
meaning that the \MgII\ line width measured from fitting with the Lorentz function is higher than that from fitting with other profiles. 
However, when we consider the full width half maximum of \MgII, we have consistent black hole mass measurements from the fitting 
with all of the four line profiles. This issue may be due to the different ratios between the FWHM and 
velocity dispersion of the line profiles. For example, the ratio is $\sim$2.35 for single Gaussian but $\sim$1 for Lorentzian. 
The black hole mass estimation method using the velocity dispersion and FWHM of \MgII\ measured from fitting 
with the line profiles of double Gaussian and Gauss-Hermite is well calibrated based on the information of \Hb\ \citep[e.g.,][]{Le2020}. 
Therefore, our test indicate that the velocity dispersion of \MgII\ from the Lorentzian fitting is not applicable for black hole mass estimation.

As described earlier, double Gaussian and Gauss-Hermite series have been widely adopted to estimate black hole mass. 
Based on our test, we found that any of them can be adopted because adopting both of them provide consistent measurements. 
However, by visual inspection, we checked Gauss-Hermite series give unsatisfactory fitting
results for a number of AGN, which fitted noise.
Therefore, we finally decided to adopt the fitting results based on the double Gaussian profile.}

Figure~2 shows the bolometric luminosities of the low redshift AGNs and SDSS quasars as a function of black hole mass.
Most of the AGNs in this work show an Eddington ratio ranging from 1\%\ to 100\%.
As shown in Figure~2, the low redshift AGNs are on average far less massive and luminous ($\rm log\ M_{\rm BH}$/\rm M$_{\odot}$ $\sim$ 8.0 
and log\ L$_{\rm Bol}$ $\sim$ 44.6) compared with the SDSS quasars ($\rm log\ M_{\rm BH}$/\rm M$_{\odot}$ 
$\sim$ 9.0 and log\ L$_{\rm Bol}$ $\sim$ 46.5).
Especially, the low redshift AGN sample uniquely covers the low luminosity range of 
log L$_{\rm Bol}$=42.1--45.4, which is $\sim$10$^{1-4}$ times fainter than the lower bound of the luminosity range of high redshift SDSS quasars 
and the low black hole mass (log M$_{\rm BH}$/\rm M$_{\odot}$= 6.9--8.0) regime. 
Together with the SDSS quasars, we can investigate the relations between the \FeII/\MgII\ ratio and AGN properties 
in the wide dynamic ranges of the AGN properties (i.e., $\sim$6 dex in luminosity and 3 dex in black hole mass).

Note that, the distributions of black hole mass and AGN bolometric luminosity of our SDSS quasars are extended by $\sim$0.5 dex to a lower mass and
luminosity, compared with those presented in Paper I. This is due to the inclusion of lower redshift SDSS quasars (0.46 < z < 0.75), which
are likely less massive and luminous than higher redshift SDSS quasars.\\

 \begin{figure}{}
\includegraphics[width = 0.44\textwidth]{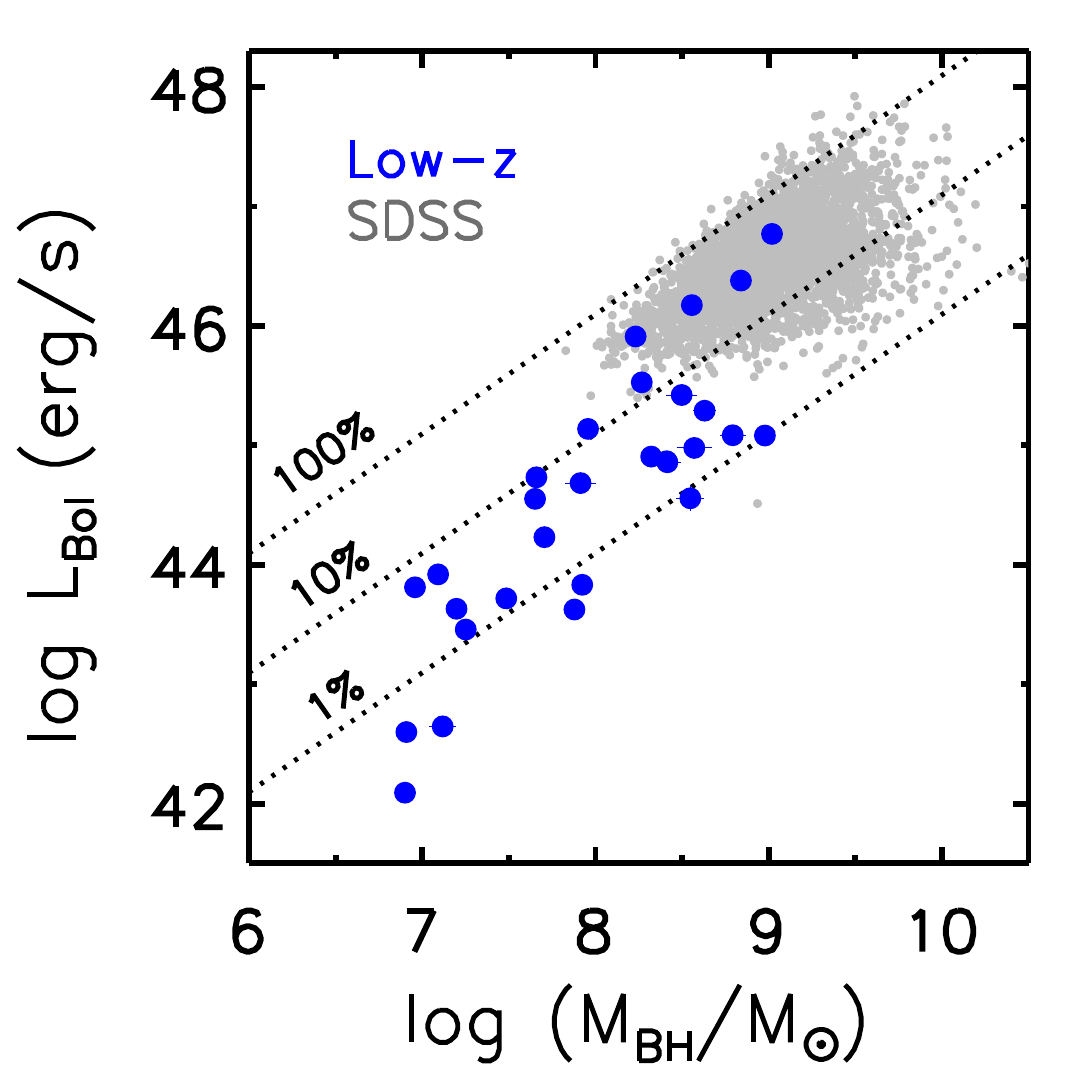}
\caption{
AGN bolometric luminosity as a function of black hole mass. Symbols are same as those in Figure~1. 
Dashed lines indicate the Eddington ratios (100, 10, and 1\%).\\
\label{fig:allspec1}}
\end{figure}

\begin{deluxetable*}{lrrcrccc} 
\tablewidth{0pt}
\tablecolumns{8}
\tabletypesize{\scriptsize}
\tablecaption{Emission line and AGN properties}

\tablehead{
\colhead{Object} &
\colhead{$\rm F_{\FeII}$} &
\colhead{$\rm F_{\MgII}$} &
\colhead{$\rm  \FeII/\MgII$} &
\colhead{$\sigma_{\rm \MgII}$}&
\colhead{log $\rm  M_{BH}/M_{\odot}$} &
\colhead{log $\rm  L_{3000}$} &
\colhead{log $\rm  L_{Bol}/L_{Edd}$} 
\\
&
 \multicolumn{2}{c}{($\rm 10^{-16}\  erg \ s^{-1} \ cm^{-2}$)}&
&
\colhead{(km s$^{-1}$)} &
&
\colhead{($\rm erg \ s^{-1}$)} &
\\
\colhead{(1)} &
\colhead{(2)} &
\colhead{(3)} &
\colhead{(4)} &
\colhead{(5)} &
\colhead{(6)} &
\colhead{(7)} &
\colhead{(8)}
}

\startdata
                     KUG 0003+199 	&$	145.8	\pm	4.6	$&$	55.3	\pm	1.0	$&$	2.64	\pm	0.10	$&$	1045	\pm	42	$&$	7.66	\pm	0.03	$&$	44.02	\pm	0.00	$&$	-1.02	\pm	0.04	$\\
              FIRST J0012-1022 	&$	10.3	\pm	2.0	$&$	3.9	\pm	0.6	$&$	2.65	\pm	0.70	$&$	1846	\pm	808	$&$	8.50	\pm	0.09	$&$	44.71	\pm	0.01	$&$	-1.17	\pm	0.15	$\\
                     PGC 87392 	&$	31.5	\pm	0.5	$&$	7.2	\pm	0.1	$&$	4.36	\pm	0.11	$&$	793	\pm	26	$&$	6.96	\pm	0.03	$&$	43.10	\pm	0.00	$&$	-1.24	\pm	0.03	$\\
                    ESO 113-45 	&$	160.6	\pm	7.1	$&$	104.2	\pm	2.0	$&$	1.54	\pm	0.09	$&$	2029	\pm	116	$&$	8.32	\pm	0.05	$&$	44.19	\pm	0.01	$&$	-1.51	\pm	0.06	$\\
                  HE 0132-4313 	&$	18.4	\pm	0.8	$&$	3.5	\pm	0.1	$&$	5.23	\pm	0.35	$&$	1025	\pm	77	$&$	8.23	\pm	0.05	$&$	45.20	\pm	0.00	$&$	-0.42	\pm	0.06	$\\
                       3C 48.0 	&$	14.7	\pm	2.5	$&$	4.2	\pm	0.3	$&$	3.54	\pm	0.30	$&$	1281	\pm	70	$&$	8.56	\pm	0.05	$&$	45.46	\pm	0.01	$&$	-0.48	\pm	0.05	$\\
      SDSS J015530.02-085704.0 	&$	11.0	\pm	1.9	$&$	7.4	\pm	0.5	$&$	1.48	\pm	0.30	$&$	2320	\pm	180	$&$	8.63	\pm	0.07	$&$	44.58	\pm	0.01	$&$	-1.44	\pm	0.06	$\\
                       AKN 120 	&$	232.3	\pm	7.8	$&$	97.7	\pm	1.8	$&$	2.38	\pm	0.11	$&$	2300	\pm	97	$&$	8.41	\pm	0.03	$&$	44.16	\pm	0.00	$&$	-1.64	\pm	0.04	$\\
                   PG 0844+349 	&$	85.6	\pm	4.1	$&$	19.1	\pm	0.5	$&$	4.48	\pm	0.29	$&$	1167	\pm	26	$&$	7.96	\pm	0.02	$&$	44.43	\pm	0.00	$&$	-0.92	\pm	0.02	$\\
                  KUG 0921+525 	&$	129.5	\pm	3.7	$&$	70.9	\pm	3.1	$&$	1.83	\pm	0.10	$&$	1150	\pm	36	$&$	7.65	\pm	0.03	$&$	43.84	\pm	0.00	$&$	-1.20	\pm	0.03	$\\
      SDSS J094603.94+013923.6 	&$	7.3	\pm	0.9	$&$	2.4	\pm	0.2	$&$	3.03	\pm	0.47	$&$	1442	\pm	155	$&$	7.91	\pm	0.09	$&$	43.97	\pm	0.03	$&$	-1.33	\pm	0.09	$\\
                      NGC 3516 	&$	184.1	\pm	7.6	$&$	122.4	\pm	2.1	$&$	1.50	\pm	0.08	$&$	2381	\pm	100	$&$	7.92	\pm	0.02	$&$	43.12	\pm	0.00	$&$	-2.19	\pm	0.04	$\\
                 MCG 10-16-111 	&$	11.1	\pm	0.7	$&$	4.6	\pm	0.2	$&$	2.44	\pm	0.19	$&$	1362	\pm	117	$&$	7.25	\pm	0.05	$&$	42.75	\pm	0.01	$&$	-1.89	\pm	0.06	$\\
                  MCG 9-19-073 	&$	17.4	\pm	1.2	$&$	16.0	\pm	0.9	$&$	1.09	\pm	0.13	$&$	1864	\pm	152	$&$	7.12	\pm	0.08	$&$	41.94	\pm	0.03	$&$	-2.57	\pm	0.07	$\\
                    CGCG 13-24 	&$	11.2	\pm	0.7	$&$	7.8	\pm	0.4	$&$	1.44	\pm	0.12	$&$	1504	\pm	99	$&$	6.91	\pm	0.06	$&$	41.89	\pm	0.03	$&$	-2.41	\pm	0.07	$\\
MARK 50	&$	37.2	\pm	2.5	$&$	17.8	\pm	0.6	$&$	2.08	\pm	0.16	$&$	1534	\pm	162	$&$	7.49	\pm	0.03	$&$	43.01	\pm	0.01	$&$	-1.86	\pm	0.06	$\\
                        3C 273 	&$	407.8	\pm	10.2	$&$	135.8	\pm	7.5	$&$	3.00	\pm	0.19	$&$	1548	\pm	77	$&$	9.02	\pm	0.04	$&$	46.06	\pm	0.00	$&$	-0.35	\pm	0.04	$\\
               RXS J12517+2404 	&$	3.8	\pm	3.5	$&$	5.5	\pm	0.4	$&$	0.69	\pm	0.45	$&$	3147	\pm	335	$&$	8.79	\pm	0.08	$&$	44.37	\pm	0.04	$&$	-1.80	\pm	0.10	$\\
                      3C 277.1 	&$	6.3	\pm	0.6	$&$	2.4	\pm	0.2	$&$	2.56	\pm	0.19	$&$	1332	\pm	274	$&$	8.27	\pm	0.04	$&$	44.82	\pm	0.01	$&$	-0.84	\pm	0.09	$\\
NGC 5548	&$	92.8	\pm	4.5	$&$	74.0	\pm	1.3	$&$	1.26	\pm	0.07	$&$	2546	\pm	103	$&$	7.88	\pm	0.02	$&$	42.91	\pm	0.01	$&$	-2.35	\pm	0.04	$\\
                     UGC 10120 	&$	32.2	\pm	0.5	$&$	11.1	\pm	0.8	$&$	2.90	\pm	0.21	$&$	869	\pm	35	$&$	7.09	\pm	0.03	$&$	43.21	\pm	0.00	$&$	-1.27	\pm	0.04	$\\
                  SBS 1701+610 	&$	3.2	\pm	8.5	$&$	4.5	\pm	0.5	$&$	0.71	\pm	0.59	$&$	3217	\pm	524	$&$	8.55	\pm	0.08	$&$	43.85	\pm	0.11	$&$	-2.09	\pm	0.13	$\\
      SDSS J171448.50+332738.3 	&$	5.6	\pm	3.0	$&$	4.4	\pm	0.3	$&$	1.27	\pm	0.35	$&$	2322	\pm	236	$&$	8.42	\pm	0.08	$&$	44.15	\pm	0.07	$&$	-1.65	\pm	0.09	$\\
                   CGCG 229-15 	&$	29.7	\pm	0.7	$&$	11.9	\pm	0.6	$&$	2.49	\pm	0.14	$&$	1159	\pm	63	$&$	7.20	\pm	0.04	$&$	42.92	\pm	0.00	$&$	-1.66	\pm	0.05	$\\
NGC 6814	&$	23.7	\pm	1.8	$&$	12.7	\pm	0.6	$&$	1.88	\pm	0.18	$&$	1996	\pm	147	$&$	6.90	\pm	0.06	$&$	41.38	\pm	0.02	$&$	-2.91	\pm	0.06	$\\
                      MARK 509 	&$	201.7	\pm	22.4	$&$	174.5	\pm	6.1	$&$	1.16	\pm	0.05	$&$	2591	\pm	368	$&$	8.57	\pm	0.10	$&$	44.27	\pm	0.01	$&$	-1.69	\pm	0.26	$\\
                   B2 2201+31A 	&$	57.4	\pm	3.5	$&$	14.9	\pm	0.5	$&$	3.85	\pm	0.30	$&$	1577	\pm	49	$&$	8.84	\pm	0.03	$&$	45.67	\pm	0.01	$&$	-0.56	\pm	0.03	$\\
                   MR 2251-178 	&$	57.0	\pm	3.0	$&$	84.9	\pm	0.9	$&$	0.67	\pm	0.04	$&$	3903	\pm	44	$&$	8.98	\pm	0.01	$&$	44.37	\pm	0.00	$&$	-1.99	\pm	0.01	$\\
                      NGC 7469 	&$	236.3	\pm	5.8	$&$	107.1	\pm	1.2	$&$	2.21	\pm	0.06	$&$	1471	\pm	25	$&$	7.71	\pm	0.01	$&$	43.52	\pm	0.00	$&$	-1.57	\pm	0.02	$
                                        \enddata
\label{table:prop}
\end{deluxetable*}

 \begin{figure}{}
\includegraphics[width = 0.46\textwidth]{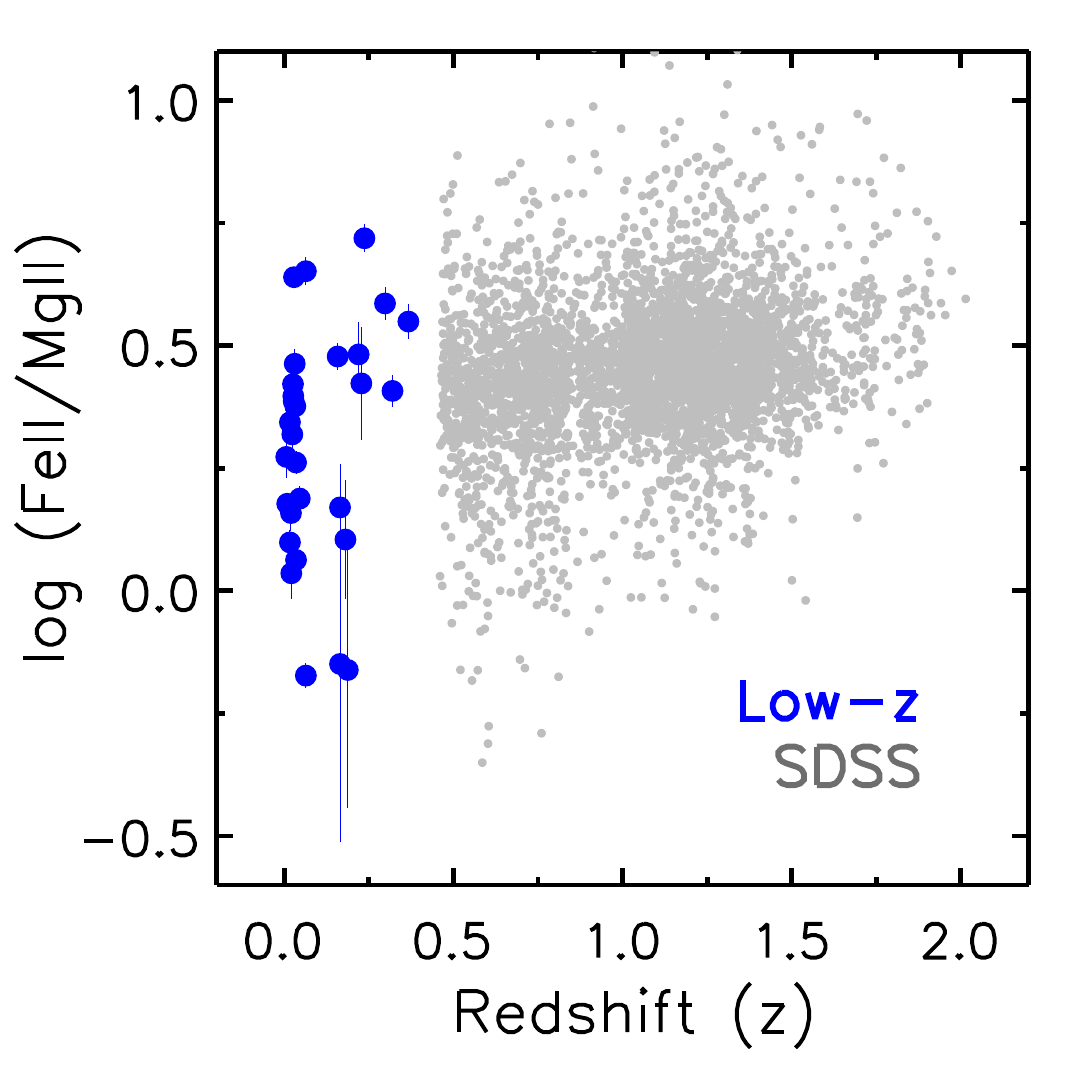}
\caption{
\FeII/\MgII\ ratio as a function of redshift.
Symbols are the same as those in Figure~1.\\
\label{fig:allspec1}}
\end{figure}

 \begin{figure*}{}
\includegraphics[width = 0.94\textwidth]{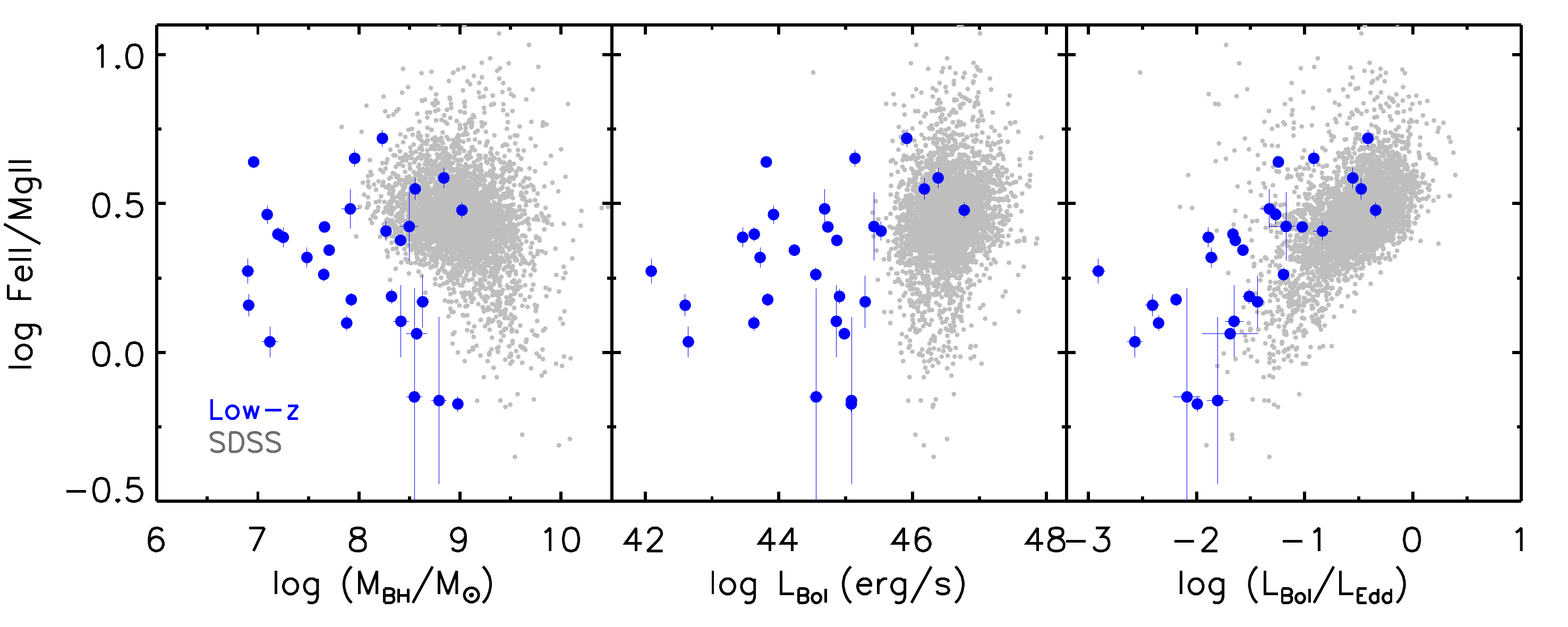}
\caption{
Relations between the \FeII/\MgII\ ratio and black hole mass (left), bolometric luminosity (middle), and Eddington ratio (right). 
Symbols are same as those in Figure~1.\\
\label{fig:allspec1}}
\end{figure*}

\section{Result} \label{section:intro}
In Figure~3, we present the \FeII/\MgII\ ratios of the low redshift AGNs and SDSS quasars as a function of redshift. 
We find that there is no significant difference in \FeII/\MgII\ ratio between the low redshift AGNs (log \FeII/\MgII\ = 0.30$\pm$0.24) and 
SDSS quasars (log \FeII/\MgII\ = 0.45$\pm$0.15). This may indicate the lack of redshift dependence of \FeII/\MgII, which is consistent 
with the findings in previous works (e.g., \citealt{Maiolino2003,DeRosa2011,Sameshima2017}; Paper I; 
\citealt{Onoue2020,Sameshima2020,Schindler2020}).

Comparisons of the \FeII/\MgII\ ratios and AGN properties are presented in Figure~4. 
We find that the \FeII/\MgII\ ratio correlates with AGN luminosity and Eddington ratio. 
Although the strong correlation between the \FeII/\MgII\ ratio and Eddington ratio has previously been reported
(\citealt{Dong2011,Sameshima2017}; Paper I), the correlation between the \FeII/\MgII\ ratio and AGN luminosity has not yet been discovered.
Meanwhile, the \FeII/\MgII\ ratio and black hole mass shows { weak (SDSS quasars) or marginal (low-z AGNs)} correlations, 
which is inconsistent with their anti-correlation found for other samples of SDSS quasars (e.g., \citealt{Dong2011}; Paper I).

For statistical assessment, we perform Spearman's { and Kendall's correlation tests for both of the low redshift AGNs and 
SDSS quasars (Table~3 and 4)}. As expected in Figure~4, we statistically confirm the strong (weak) correlation of the \FeII/\MgII\ ratio 
with Eddington ratio (AGN luminosity). In addition, the correlation coefficients are found to be larger for 
the low redshift AGNs than for the SDSS quasars. In terms of the relation between the \FeII/\MgII\ and black hole mass, 
we find no correlation for the low redshift AGNs, but a weak anti-correlation for the SDSS quasars. 
{ Overall, the statistical tests may indicate that the \FeII/\MgII\ ratio strongly correlates with Eddington ratio rather than
black hole mass and luminosity. In the appendix, we provide the fitting results of 5 SDSS quasars with low- and high- Eddington ratio, 
respectively, to show how their spectra are different depending on Eddington ratio.}
{ We find that the correlations between the \FeII/\MgII\ ratio and AGN properties shown for the SDSS quasars 
are consistent with those in the literature (e.g., \citealt{Dong2011}; Paper I). Note that the AGN properties presented 
in \cite{Dong2011} were calculated based on the line width of \Hb\ and 5100\AA\ continuum luminosity. 
This means that the choice of emission lines between \Hb\ and \MgII\ for black 
hole mass estimation does not affect the main result.}
\\

\begin{deluxetable}{cccc} 
\tablewidth{0pt}
\tablecolumns{8}
\tabletypesize{\scriptsize}
\tablecaption{Results of Spearman's Rank-order correlation test}
\tablehead{
&
\colhead{$M_{\rm BH}/M_{\odot}$} &
\colhead{$L_{\rm Bol}/L_\odot$} &
\colhead{$L_{\rm Bol}/L_{\rm Edd}$} 
\\
\colhead{(1)} &
\colhead{(2)} &
\colhead{(3)} &
\colhead{(4)}  }
\startdata
\FeII/\MgII\  (low-z) &$r_{s}$= --0.121 & $r_{s}$= 0.342 & $r_{s}$= 0.776 	\\
	&$p$=  0.531  &$p$= 0.069 &$p$= 0.000         \\
\hline
\FeII/\MgII\ (SDSS quasars) &$r_{s}$= --0.250 & $r_{s}$= 0.192 & $r_{s}$= 0.502 	\\
	&$p$=  0.000  &$p$= 0.000 &$p$= 0.000         
\enddata
\label{table:prop}
%\tablecomments{
%  }
\end{deluxetable}

\begin{deluxetable}{cccc} 
\tablewidth{0pt}
\tablecolumns{8}
\tabletypesize{\scriptsize}
\tablecaption{Results of Kendall's Rank-order correlation test}
\tablehead{
&
\colhead{$M_{\rm BH}/M_{\odot}$} &
\colhead{$L_{\rm Bol}/L_\odot$} &
\colhead{$L_{\rm Bol}/L_{\rm Edd}$} 
\\
\colhead{(1)} &
\colhead{(2)} &
\colhead{(3)} &
\colhead{(4)}  }
\startdata
\FeII/\MgII\  (low-z) &$r_{s}$= --0.099 & $r_{s}$= 0.212 & $r_{s}$= 0.557 	\\
	&$p$=  0.453  &$p$= 0.101 &$p$= 0.000         \\
\hline
\FeII/\MgII\ (SDSS quasars) &$r_{s}$= --0.172 & $r_{s}$= 0.130 & $r_{s}$= 0.361 	\\
	&$p$=  0.000  &$p$= 0.000 &$p$= 0.000         
\enddata
\label{table:prop}
\end{deluxetable}

\section{Discussion} \label{section:intro}

\begin{figure}{}
\includegraphics[width = 0.44\textwidth]{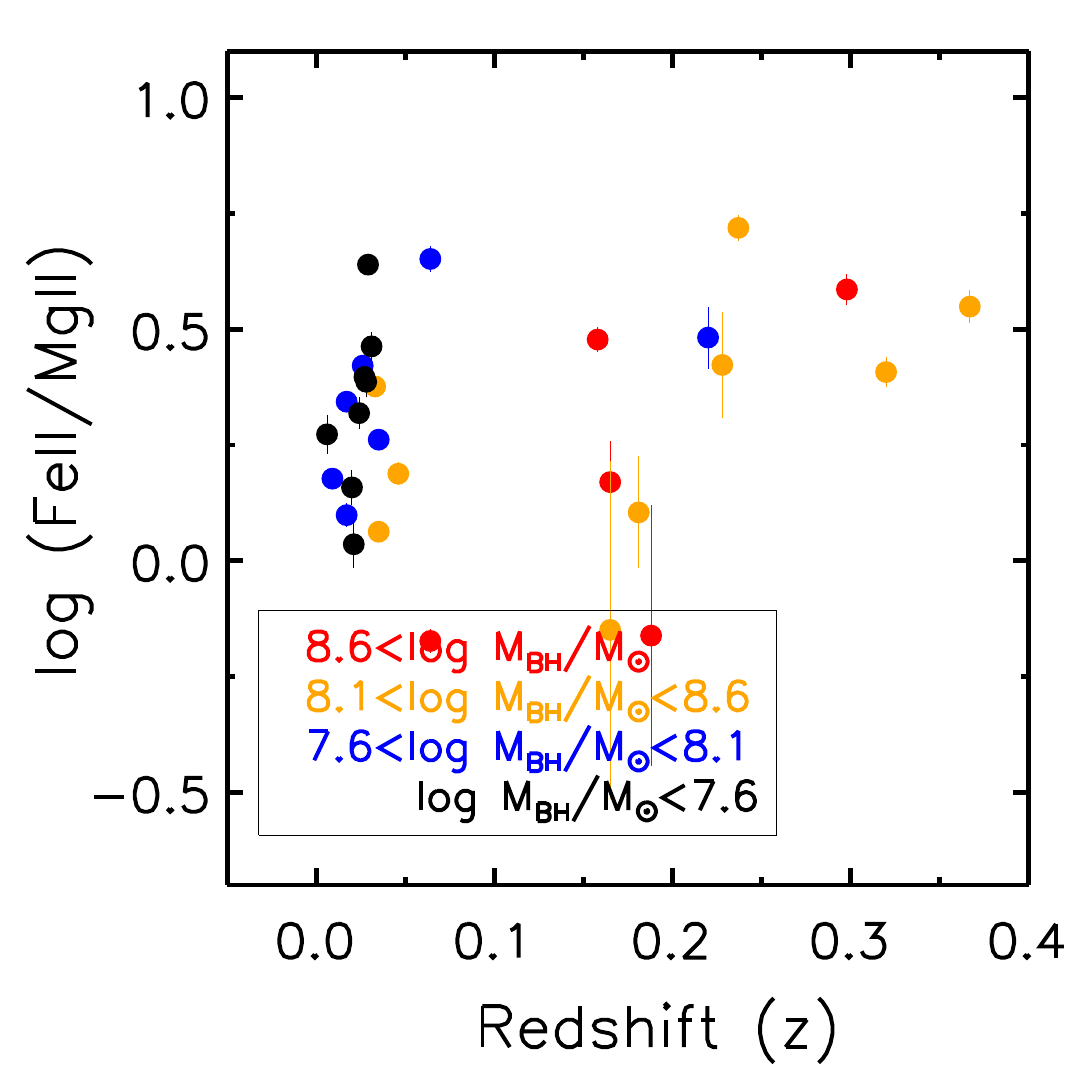}
\caption{
\FeII/\MgII\ ratios of the low redshift AGNs as a function of redshift. Color represents black hole mass bins.\\
\label{fig:allspec1}}
\end{figure} 

\subsection{Lack of redshift dependence of \FeII/\MgII} \label{section:intro}
Extending the redshift range down to $\sim$0, we confirm that the \FeII/\MgII\ ratio is independent of redshift up to z$\sim$7
\citep{Barth2003,Maiolino2003,DeRosa2011,Mazzucchelli2017,Shin2019,Onoue2020,Sameshima2020,Schindler2020}.
However, in Paper I, we discussed that the absence of cosmic evolution of the \FeII/\MgII\ ratio may be caused 
by biased samples in previous works, which investigated very luminous and massive AGNs.
The black hole mass range of the SDSS quasars presented in Paper I is $\sim$ 10$^{8-10} M_{\odot}$. 
With the assumption of a black hole mass--stellar mass ratio of 0.001--0.01 
\citep[e.g.,][]{Marconi2003,Peng2006,Schramm2008}, the corresponding mass of their host galaxies 
ranges 10$^{10-13} M_{\odot}$.  At such high stellar mass regimes, galaxy metallicity evolution is known 
to be non-significant as a function of redshift \citep[e.g.,][]{Erb2006,Maiolino2008}. 
For example, for galaxies with a stellar mass of 10$^{11} M_{\odot}$, metallicity increases by only $\sim$0.2 dex 
between redshifts 0.4 and 2, which is the range of our SDSS quasars (see Figures~8 and 9 of \citealt{Maiolino2008}). 
Therefore, even if the cosmic evolution of the \FeII/\MgII\ ratio does exists, it may be difficult to detect the evolution 
due to the large scatter of the \FeII/\MgII\ ratio.

To better understand this issue, we investigate the low redshift AGNs since their black hole masses are down to 
{ $\sim$ 10$^{6.9} M_{\odot}$}, corresponding to a mass of their host galaxies of { 10$^{8.9-9.9} M_{\odot}$}  
\citep[see][and the references therein]{Greene2020}. As shown in Figure~3, the low redshift AGNs do not show any clear 
redshift evolution of the \FeII/\MgII\ ratio, whereas the black hole masses of the low redshift AGNs span { two} 
orders of magnitude. Since galaxy metallicity-redshift relations depend on stellar mass \citep[see e.g.,][]{Maiolino2008},   
we may need to take into account a black hole mass dependency in the \FeII/\MgII-redshift relation.
To this end, we divide the low redshift AGN sample into four groups on the basis of their black hole masses 
and investigate the \FeII/\MgII--redshift relation for each subsample (Figure~5). 
However, we find no significant evidence for the redshift evolution of the \FeII/\MgII\ ratio across all the
black hole mass bins. Although this result potentially confirms the absence of cosmic evolution in the \FeII/\MgII\ ratio, 
it could have been due to the narrow redshift range of the low redshift AGN sample (i.e., z < $\sim$0.4) and 
the large scatter of \FeII/\MgII. It has been observed that galaxy metallicity increases
by less than $\sim$0.2 dex within the redshift range of 0-0.7 \citep[see e.g., Figure~9 of ][]{Maiolino2008}. 
With this reason, it is possible that the redshift evolution of the \FeII/\MgII\ ratio of the low redshift AGNs
is concealed by the large scatter (i.e., $\sim${0.24} dex), which is similar to that of higher redshift AGNs.\\

\subsection{Primary parameter: Eddington ratio?} \label{section:intro}
In Section 4, we confirm that \FeII/\MgII\ shows a strong dependency on Eddington ratio using a sample of low redshift AGNs 
over large dynamic ranges of black hole mass and luminosity, which is consistent with previous works 
(\citealt{Dong2011,Sameshima2017} and Paper I). However, in contrast to the previous works (\citealt{Dong2011} and Paper I), 
we find that the \FeII/\MgII\ ratio exhibits different trends in terms of black hole mass and AGN luminosity.

We reveal, for the first time, the weak positive correlation between the \FeII/\MgII\ ratio and AGN luminosity, which is
consistent with previous results for the BLR metallicity indicators \citep[i.e., \NV/\CIV,][]{Hamann1993,Nagao2006, Matsuoka2011b,Shin2013}.
It can be postulated that the correlation could have been revealed because of the wider dynamic range ($\sim$6 dex) 
of the AGN luminosity than that adopted in the previous works (i.e., $\sim$1.5 dex, e.g., \citealt{Dong2011}; Paper I).
If we take into account the large scatter of the \FeII/\MgII\ ratio (see Figure~4), it may be difficult to observe the correlation 
within a narrow dynamic range of luminosity. Our interpretation can be supported through the weaker correlation 
between the \FeII/\MgII\ ratio and AGN luminosity of the SDSS quasars, 
the luminosity range of which is $\sim$2 dex, compared with that of the low redshift AGNs (see Table~3). 

 \begin{figure*}{}
\includegraphics[width = 0.94\textwidth]{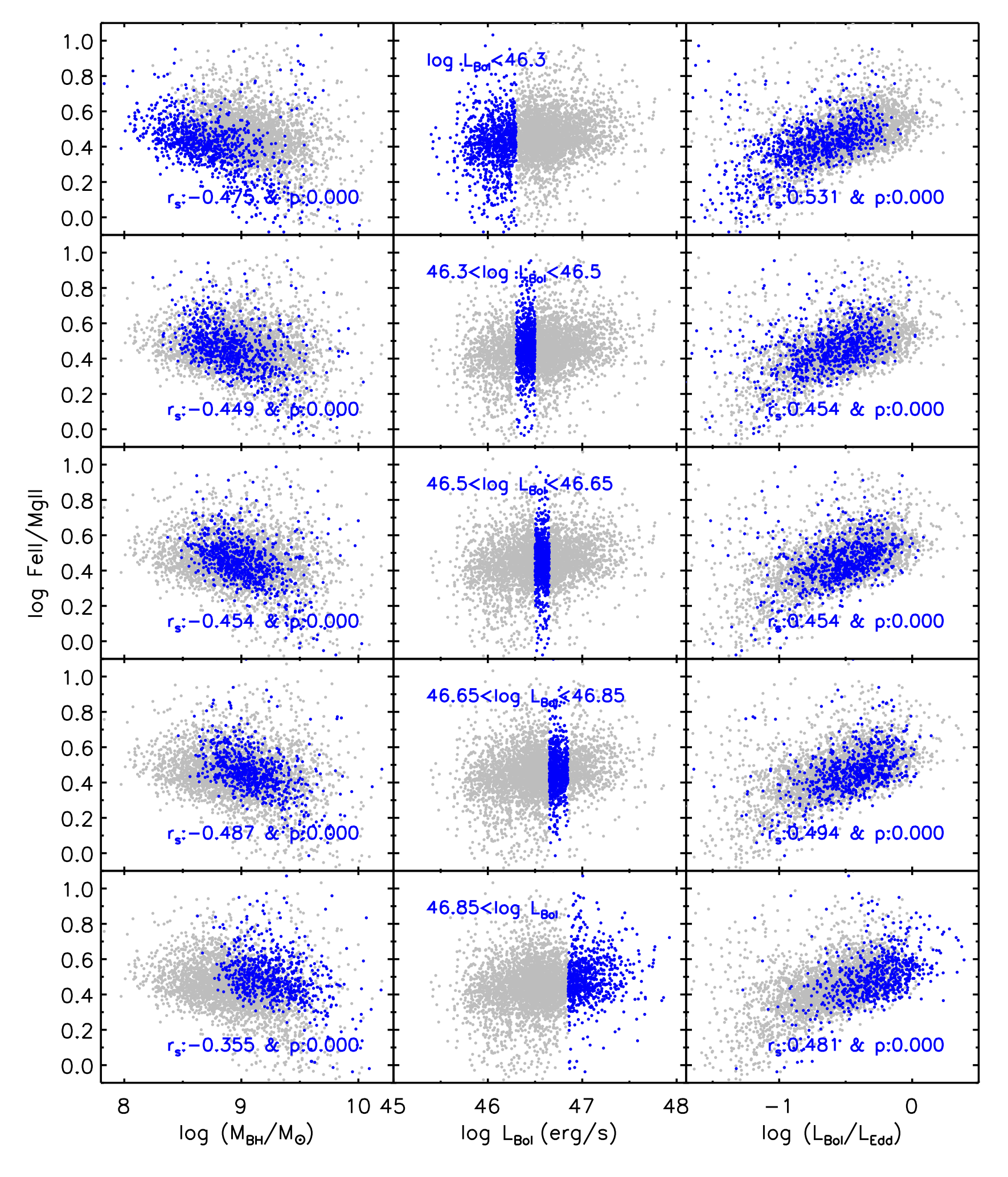}
\caption{
Relations between \FeII/\MgII\ ratio and AGN properties for the subsamples of the SDSS quasars
 in different luminosity bins, which are denoted in the left column. 
Lowest to highest redshift bins are shown from top to bottom.
Blue and gray symbols indicate a subsample and the whole sample of the SDSS quasars, respectively.
Correlation coefficients between the \FeII/\MgII\ ratio and AGN properties for each subsample are shown in
each panel.\\
\label{fig:allspec1}}
\end{figure*} 

 \begin{figure*}{}
\includegraphics[width = 0.94\textwidth]{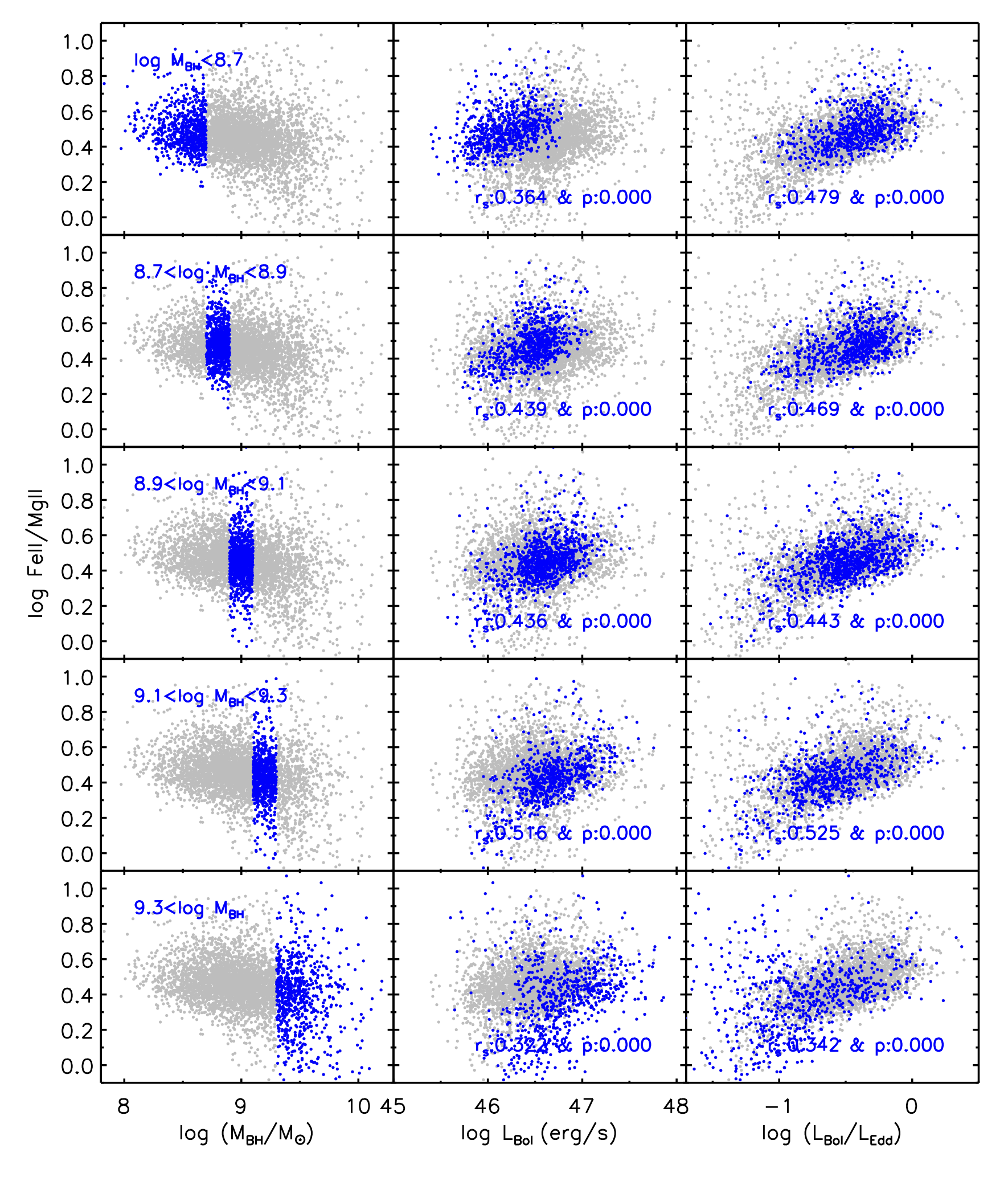}
\caption{
Same as Figure~6 except with subsamples in different black hole mass bins, which are denoted in the left column.\\
\label{fig:allspec1}}
\end{figure*} 

 \begin{figure*}{}
\includegraphics[width = 0.94\textwidth]{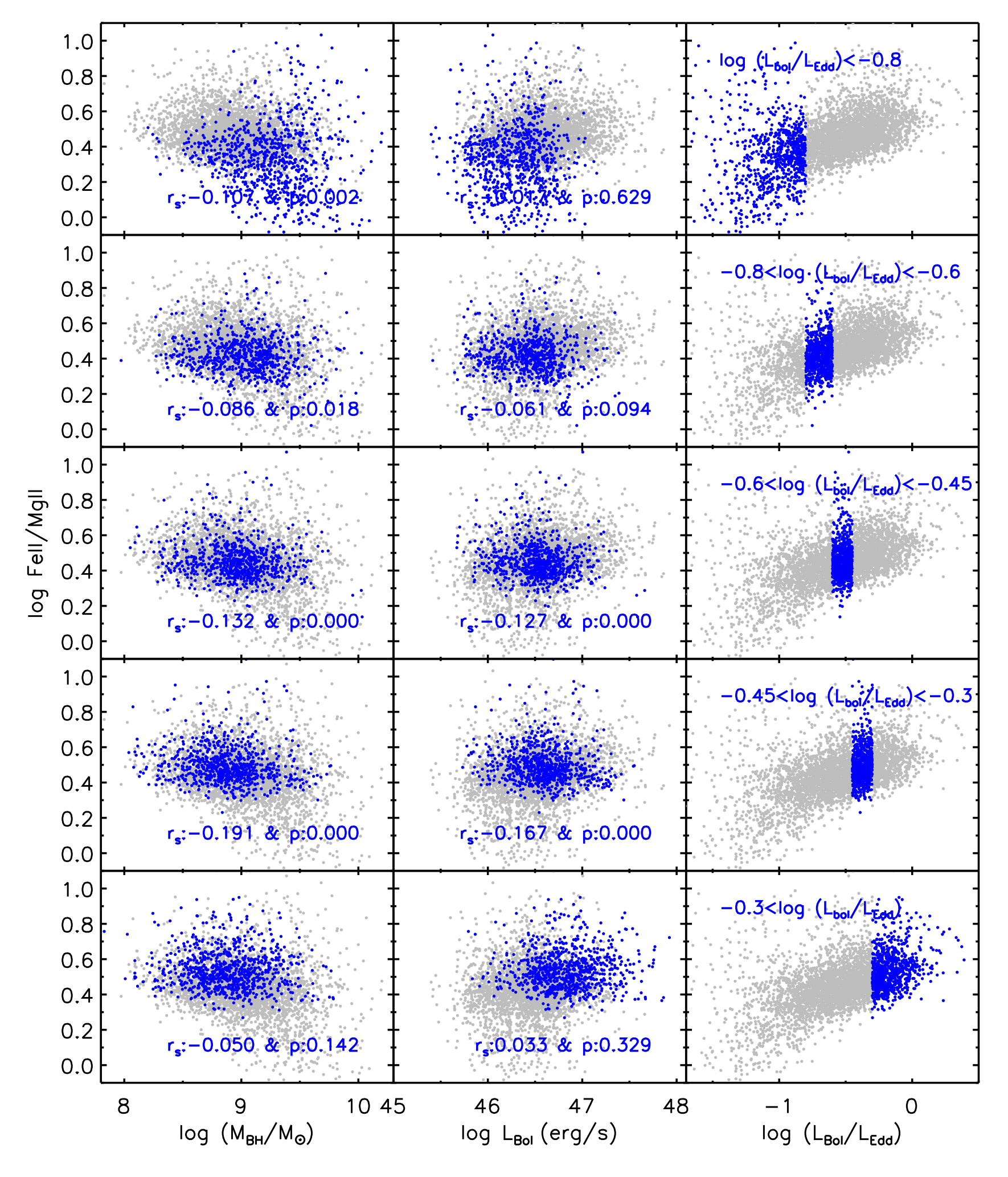}
\caption{
Same as Figure~6 except with subsamples in different Eddington ratio bins, which are denoted in the right column.\\
\label{fig:allspec1}}
\end{figure*}

{ 
To better understand the relations between the \FeII/\MgII\ ratio and AGN properties,
we divide our 4,061 SDSS quasars into 5 subsamples based on luminosity, mass, and 
Eddington ratio and investigate the relations. Note that we arbitrarily determine 
the physical ranges for each subsample to include $\sim$800 quasars. 

First, as shown in Figure~6, the \FeII/\MgII\ ratio shows 
a strong correlation (anti-correlation) with Eddington ratio (black hole mass) within 
a narrow AGN luminosity range in each luminosity bin. 
Interestingly, the correlation coefficient between the \FeII/\MgII\ ratio and black hole mass 
for each subsamples is much larger than that for the whole sample of SDSS 
quasars and low redshift AGNs. On the other hand, the correlation coefficient between the 
\FeII/\MgII\ ratio and Eddington ratio are similar for the subsamples and whole samples of SDSS quasars. 
The different behaviors in the relation between the \FeII/\MgII\ ratio and black hole mass depending on 
samples may indicate the sample selection effect. At a given luminosity, Eddington ratio and black hole mass are 
inversely proportional. Therefore, if there is a strong correlation between the \FeII/\MgII\ ratio and Eddington ratio,
it can lead the strong anti-correlation between the \FeII/\MgII\ ratio and black hole mass 
for each subsample within a narrow luminosity range. 

Second, in each mass bin, there are strong correlations of the \FeII/\MgII\ ratio with both of AGN luminosity and 
Eddington ratio (Figure~7). At a given black hole mass, AGN luminosity and Eddington ratio are proportional.
Therefore, the strong correlation between the \FeII/\MgII\ and AGN luminosity in each subsample can be also caused by 
the strong correlation between the \FeII/\MgII\ ratio and Eddington ratio.}

Third, Figure~8 shows that the \FeII/\MgII\ ratio is marginally correlated with both of black hole mass and AGN 
luminosity in each Eddington ratio bin. These results potentially confirm that the \FeII/\MgII\ ratio--black hole mass 
and \FeII/\MgII\ ratio--AGN luminosity relations are subject to sample selection, suggesting that the reported 
anti-correlation between the \FeII/\MgII\ ratio and black hole mass may be artificial. 
Meanwhile, regardless of the sample selection, Eddington ratio shows a strong correlation with the \FeII/\MgII\ ratio.

In conclusion, the \FeII/\MgII\ ratio is primarily related to Eddington ratio, 
and not black hole mass, which is related to the global properties of galaxies (i.e., galaxy mass). 
This trend is consistent with previous results for the BLR metallicity indicators \citep[e.g.,][]{Shemmer2004,Shin2013} and
other emission line ratios \citep[i.e., \FeII/\Hb, e.g.,][]{Netzer2007,Dong2011}, 
implying the physical connection between the metal enrichment at the nuclear region of galaxies and the accretion activity of AGNs 
(e.g., \citealt{Dong2011,Shin2013,Shin2017}; Paper I).\\

{
\subsection{Physical meaning of the \FeII/\MgII\ ratio} \label{section:intro}
Several line ratios (i.e., \NV/\CIV\ and \FeII/\Hb) including the \FeII/\MgII\ ratio, correlate with Eddington ratio.
Interestingly, the \FeII/\Hb\ ratio \citep{Panda2018,Panda2019a} and \NV/\CIV\ ratio 
\citep{Hamann1992,Hamann1993,Nagao2006} have been suggested to be associated with BLR metallicity.
Therefore, the correlation between the \FeII/\MgII\ ratio and Eddington ratio may suggest that the \FeII/\MgII\ ratio is also 
related to BLR metallicity, also supported by theoretical studies \citep{Verner2003,Sarkar2021}.
    
A physical connection between the \FeII/\MgII\ ratio and BLR metallicity can be understood in terms of 
metal enrichment from star formation. As a proxy of the Fe/Mg abundance ratio, we can track the 
change of the \FeII/\MgII\ ratio as a function of time from the onset of star formation based on an 
enrichment delay \citep[tens of Myr to a few Gyrs,][]{Matteucci2001} 
between alpha elements and iron which are produced mainly from Type II and Type Ia supernovae, respectively. 
If we simply consider a single episode of star formation, the \FeII/\MgII\ will be constant from the onset of star formation 
until Type II SNe explode. Once Type II SNe begin to explode (a few tens of Myrs from the onset of star formation),
the \FeII/\MgII\ will decrease due to the enrichment of alpha elements during other few tens of Myrs until
all of Type II SNe explode. After then, iron will be enriched by Type Ia SNe; hence the \FeII/\MgII\ ratio
will increase as a function of time for a few Gyrs. Therefore, except for a short time period ($\sim$a few tens of Myr), 
it can be expected that the \FeII/\MgII\ ratio increases as a function of time.

On the other hand, metallicity increases as a function of time from the onset of star formation 
due to the enrichment of metals from stellar evolution. In this sense, the possible connection 
between the \FeII/\MgII\ ratio and BLR metallicity can be considered. We note that there were 
reports that the \FeII/\MgII\ ratio is not only determined by the Fe/$\alpha$ abundance ratio but also other 
physical parameters of BLR cloud \citep[i.e., gas density and microturbulence][]{Verner2003,Sameshima2017}. 
With this reason, the \FeII/\MgII\ may not be a direct indicator of metallicity, while it can still be an 
indirect indicator of metallicity as we discussed. 

In conclusion, the correlation between several line flux ratios tracing the chemical properties of the BLR and 
Eddington ratio can be considered as the relation between the BLR metallicity and accretion activity of AGN. 
In the next section, we further discuss the origin for the relation between the BLR metallicity and Eddington ratio.}\\

\subsection{Physical implication of the relation between the BLR metallicity and Eddington ratio} \label{section:intro}

{ For the last few decades, the connection between the BLR metallicity and 
Eddington ratio has long been discussed, while it has not been fully understood. 
From a number of optical studies on AGN \citep[e.g.,][]{Boroson1992,Sulentic2000a,Sulentic2000b,Marziani2001,Boroson2002,Shen2014},
the \FeII/\Hb\ ratio and Eddington ratio are known to be related to the AGN Eigenvector 1. 
It means that the BLR metallicity and Eddington ratio can play important roles in AGN physics 
since the \FeII/\Hb\ ratio is associated with the BLR metallicity. Understanding physical connection 
between the BLR metallicity and Eddington ratio can therefore shed light on the AGN physics.

In previous works, nuclear starburst has been discussed as the origin for the relation because it
can enrich BLR metallicity as well as increase accretion rate of AGNs 
\citep[][]{Shemmer2004,Netzer2007,Shin2013,Shin2017,Panda2019b}.}
Indeed, the connection between nuclear starburst and the accretion activity of AGNs 
has been discussed \citep[e.g.,][]{Imanishi2004,Woo2012,Esquej2014} with a delay time of 
50--100 Myr between the two activities \citep{Davies2007}. However, no strong correlation was found between the 
nuclear star formation rate and Eddington ratio \citep{Davies2007,Woo2012}.
For example, in \cite{Davies2007}, Circinus galaxy, NGC 3783, and IRAS 05189-2524 
show similar nuclear star formation rates (30-70 M$_{\odot}$/yr) 
and starburst ages (i.e., 60--80 Myr), while their Eddington ratios are quiet different by factors of up to $\sim$50. 
This result suggests that star formation rate and the accretion rate of AGNs may be not directly related, meaning that the nuclear 
starburst may not be the origin for the correlation between the chemical properties in the BLR and Eddington ratio. 
We note that, however, it is also possible that the no clear relation 
between the nuclear star formation rate and Eddington ratio is due to AGN variability \citep[e.g.,][]{Hickox2014}. 

Alternatively, we can discuss metal cooling as a possible mechanism to explain the connection between 
the chemical properties in the BLR and Eddington ratio.  
Since the efficiency of radiative cooling is a function of metallicity \citep[e.g.,][]{Sutherland1993,Wang2014},
more efficient radiative cooling due to higher BLR metallicity (and also the Fe/$\alpha$ abundance ratio, 
see e.g., \citealt{Hamann1993}) can cause stronger gas inflows (i.e., higher gas density) in the BLR 
as well as higher accretion rate of AGNs.
This scenario is supported by a simulation based study \citep{Mayer2015}, finding
that the metal cooling enhances the gas inflows into the central pc scale region of galaxies.
 { We note that, based on CLOUDY photoionization modeling \citep{Ferland2017}, \cite{Sameshima2017} 
 discussed an anti-correlation between hydrogen gas density of BLR clouds and Eddington ratio, 
 which is opposite to our scenario. However, their modeled emission line strengths of \FeII\ and \MgII\ are found to be 
 largely inconsistent ($\sim$1 dex) with the observed line strengths (see Figures~4 and 9 of \citealt{Sameshima2017}), 
 which indicates that their result may not be reliable. The inconsistency can be because their adopted atomic 371-level 
 Fe$^{+}$ model \citep{Verner1999} may be not accurate \citep[see e.g.,][]{Verner2003,Smyth2019,Sarkar2021}
 and/or their parameter space may not reflect the nature of the BLR. Therefore, to confirm the relation between gas 
 density and Eddington ratio, further investigations are required.}
 \\

\section{Summary}\label{summary}
In this work, we investigate the relations between the \FeII/\MgII\ ratio and AGN properties of type 1 AGNs at z < 2 
in a large dynamic range of AGN properties. The main results are summarized below.\\

$\bullet$ The \FeII/\MgII\ ratios of the low redshift AGNs (z < $\sim$0.4) show no clear dependence on redshift, which is consistent with 
previous results for higher redshift AGNs (up to z$\sim$7), suggesting no cosmic evolution of the \FeII/\MgII\ ratios. \\

$\bullet$ In contrast to previous works, we reveal a weak positive correlation between 
the \FeII/\MgII\ ratio and AGN luminosity over a wide dynamic range of AGN luminosity ($\sim$ 6 dex).
This result is consistent with the previous studies of BLR metallicity indicators (i.e., \NV/\CIV).\\

$\bullet$ We find a marginal correlation between the \FeII/\MgII\ ratio and black hole mass, which
is inconsistent with the anti-correlation found for SDSS quasars in the literature. 
Based on a test using various subsamples of SDSS quasars, we interpret that the 
previous finding of the anti-correlation was caused by the narrow luminosity range and 
there may be no physical relation between the \FeII/\MgII\ ratio and black hole mass. \\

$\bullet$ Consistent with previous results, we find a strong correlation between the \FeII/\MgII\ ratio 
and Eddington ratio. With regard to this, we discuss metal cooling, which can enhance gas inflows 
as well as high accretion rate of AGNs. \\

\acknowledgements
{ We thank the anonymous referee for valuable comments and suggestions that significantly improved the quality of the paper. }
This research was supported by Basic Science Research Program through the National Research Foundation of Korea (NRF) funded by the Ministry of Education (2016R1A2B3011457 and 2019R1A6A3A01093189). JS and MK were supported by the National Research Foundation of Korea (NRF) grant funded by the Korea government (MSIT) (No. 2020R1A2C4001753). TN was financially supported by the Japan Society for the Promotion of
Science (JSPS) KAKENHI grant Nos. 19H00697 and 20H01949.

Based on observations made with the NASA/ESA Hubble Space Telescope, and obtained from the Hubble Legacy Archive, which is a collaboration between the Space Telescope Science Institute (STScI/NASA), the Space Telescope European Coordinating Facility (ST-ECF/ESAC/ESA) and the Canadian Astronomy Data Centre (CADC/NRC/CSA).

Funding for the SDSS and SDSS-II has been provided by the Alfred P. Sloan Foundation, the Participating Institutions, the National Science Foundation, the U.S. Department of Energy, the National Aeronautics and Space Administration, the Japanese Monbukagakusho, the Max Planck Society, and the Higher Education Funding Council for England. The SDSS Web Site is http://www.sdss.org/.

The SDSS is managed by the Astrophysical Research Consortium for the Participating Institutions. The Participating Institutions are the American Museum of Natural History, Astrophysical Institute Potsdam, University of Basel, University of Cambridge, Case Western Reserve University, University of Chicago, Drexel University, Fermilab, the Institute for Advanced Study, the Japan Participation Group, Johns Hopkins University, the Joint Institute for Nuclear Astrophysics, the Kavli Institute for Particle Astrophysics and Cosmology, the Korean Scientist Group, the Chinese Academy of Sciences (LAMOST), Los Alamos National Laboratory, the Max-Planck-Institute for Astronomy (MPIA), the Max-Planck-Institute for Astrophysics (MPA), New Mexico State University, Ohio State University, University of Pittsburgh, University of Portsmouth, Princeton University, the United States Naval Observatory, and the University of Washington.\\

%\acknowledgements

{

%\restartappendixnumbering
\appendix
In Figure~A1, we provide fitting results of five randomly selected SDSS quasars 
with low (log L$_{\rm Bol}$/L$_{\rm Edd}$<-1) and high Eddington ratio (log L$_{\rm Bol}$/L$_{\rm Edd}$>0), 
respectively (Figure~A1). As expected from Figure~4, the high Eddington ratio quasars show 
larger \FeII/\MgII\ ratios than the low Eddington ratio quasars. We find one difference between the two groups that
low Eddington ratio AGNs show much broader \MgII\ line than high Eddington ratio AGNs.
This trend can be expected from i) a strong anti-correlation between the line width of \Hb\ and
Eddington ratio \cite{Netzer2007,Xu2012} and ii) a linear relationship between
the linewidths of \Hb\ and \MgII\ \citep[e.g.,][]{Mcgill2008,Woo2018,Le2020}.
This trend also can be analytically explained. Based on the single epoch method, 
black hole mass is proportional to L$^{\sim0.5}\times\ \sigma^{\sim2}$.
Since Eddington ratio is determined by the ratio between bolometric luminosity
and black hole mass, it is proportional to L$^{\sim0.5}\times\ \sigma^{\sim-2}$, which means 
a strong inverse dependency of Eddington ratio on the line width of \Hb\ or \MgII.}\\

\renewcommand{\thefigure}{A\arabic{figure}}
\addtocounter{figure}{-8}

 \begin{figure}{}
\hspace*{-0.55cm} 
\includegraphics[width = 0.51\textwidth]{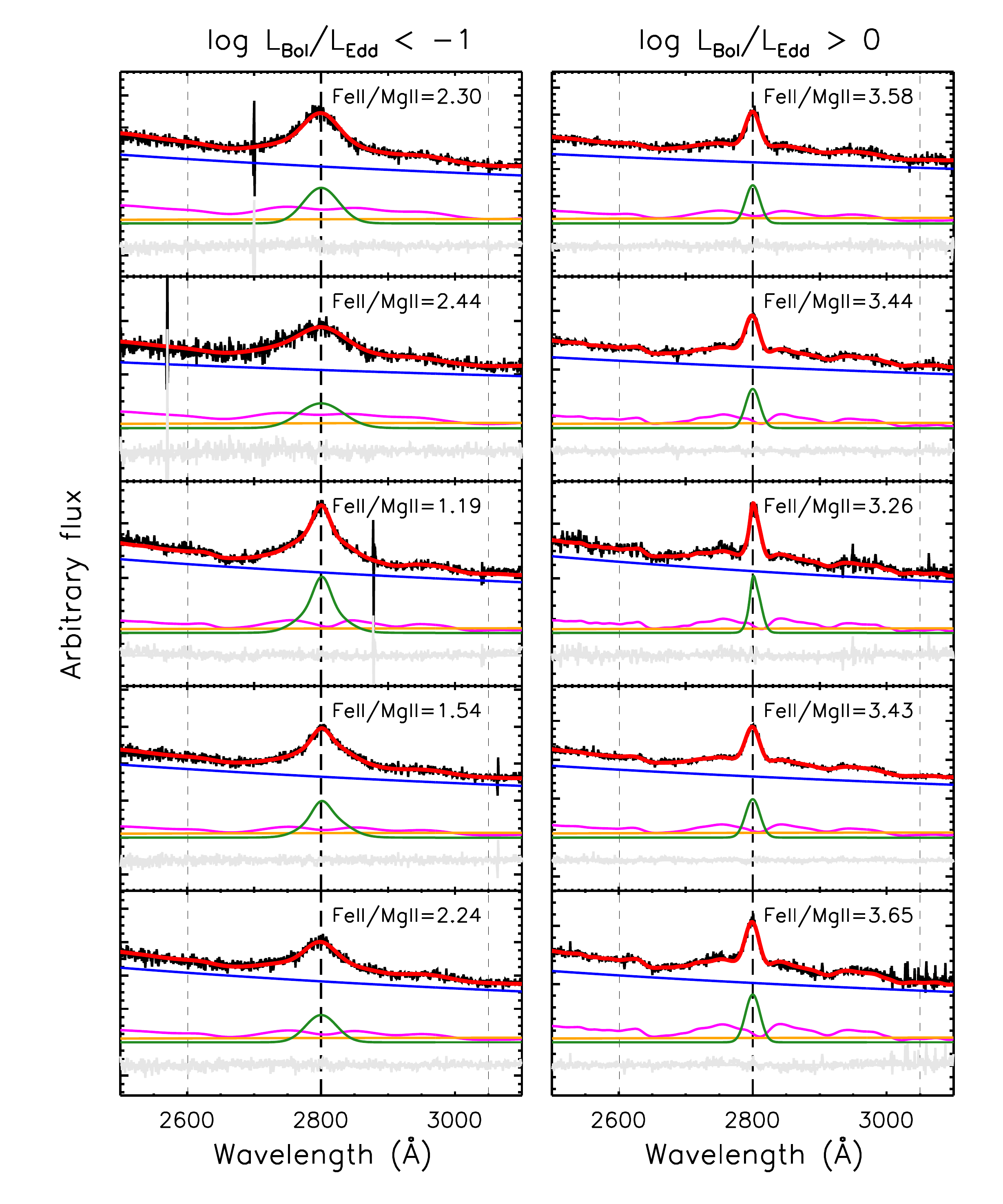}
\caption{
{ Examples of fitting results of SDSS quasars with low- (left) and high- Eddington ratio (right). In each panel,
raw spectrum (black), the best-fit model (red), power-law continuum (blue), Balmer pseudo continuum (orange),
\FeII\ (magenta), \MgII\ (green), and the residual spectrum (gray) are shown. The \FeII/\MgII\ ratio of each quasar is
also shown at the top right corner of each panel.\\}
\label{fig:allspec1}}
\end{figure}

%\bibliography{feii_mgii}
%\clearpage

\end{document}